\documentclass[12pt]{article}

\usepackage{times}
\usepackage{graphicx}
\usepackage{psfrag}
\usepackage{amsmath,amsthm}
\usepackage{amsfonts}
\usepackage{color}
\usepackage{authblk}
\usepackage{mychicago}
\usepackage{genetics_manu_style}

\newcommand{\var}{\operatorname{Var} }
\usepackage{setspace}
\begin{document}

\parskip=10pt

\title{Population genetics models of local ancestry} 
\author{Simon Gravel}
\affil{Genetics Department\\Stanford University\\Stanford, CA, 94305-5120}


\renewcommand{\CorrespondingAddress}{Department of Genetics\\
Stanford University\\
300 Pasteur Dr.\\
Lane Building, Room L337\\
Stanford, CA, 94305-5120\\
Phone: 650-644-7625\\
Fax: 650-723-3667\\
 \vfill}
\renewcommand{\RunningHead}{Genetic models of local ancestry}
\renewcommand{\CorrespondingAuthor}{Simon Gravel}
\renewcommand{\KeyWords}{Admixture, Local ancestry, Demographic inference, Population structure, Gene flow }

\maketitle




\pagebreak

\begin{abstract}

Migrations have played an important role in shaping the genetic diversity of human populations. Understanding genomic data thus requires careful modeling of historical gene flow. Here we consider the effect of relatively recent population structure and gene flow, and interpret genomes of individuals that have ancestry from multiple source populations as mosaics of segments originating from each population. This article describes general and tractable models for local ancestry patterns with a focus on the length distribution of continuous ancestry tracts, and the variance in total ancestry proportions among individuals. The models offer improved agreement with Wright-Fisher simulation data when compared to the state-of-the art, and can be used to infer time-dependent migration rates from multiple populations. Considering HapMap African-American (ASW) data, we find that a model with two distinct phases of `European' gene flow significantly improves the modeling of both tract lengths and ancestry variances.


\end{abstract}

\section{Introduction}
DNA sequencing is an invaluable tool for understanding demographic relationships between populations. Even with a limited number of genetic markers, measured across individuals and populations, it is often possible to estimate relatedness between populations, ancestry proportions in admixed populations, or sex-biased gene flow. The availability of dense genotyping platforms and high-throughput sequencing technology has enabled refined analyses of genetic diversity. 

\begin{figure}
\scalebox{.6}{\includegraphics{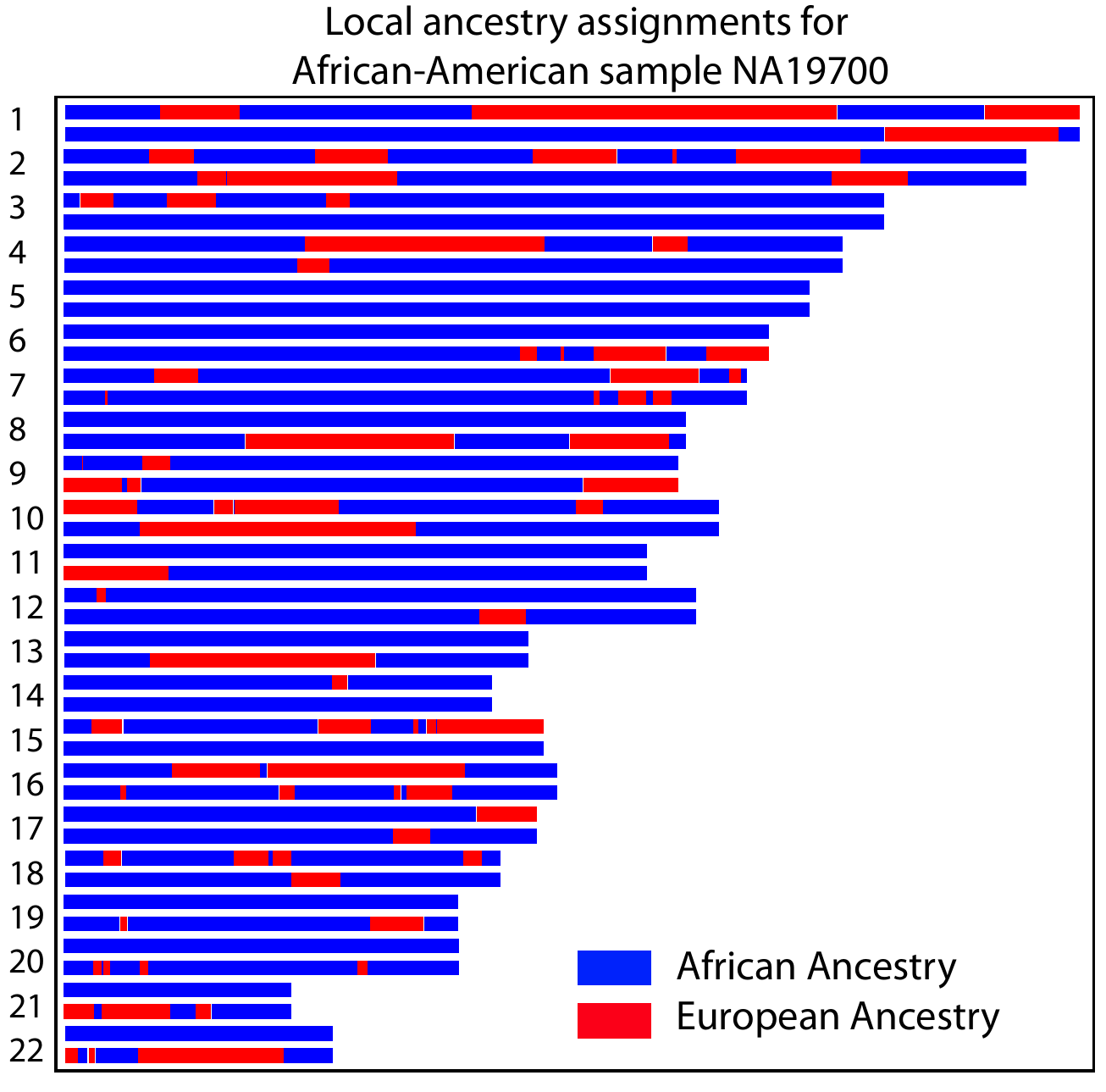}}
\caption[karyogram]{\label{karyogram}Local ancestry across 22 autosomes for an African American individual inferred by PCAdmix, a local ancestry inference software \protect \cite{abra_s_thesis} using HapMap European (CEU) and Yoruba (YRI) as source populations. The majority of the genome is inferred to be of African origin (blue), but a significant fraction of the genome is inferred to be of European origin (red). The purpose of this article is to model the distribution of ancestry assignments in such admixed individuals.}
\end{figure}

Because of recombination, different loci along an individual genome can reveal different aspects of its ancestry. Consider a sample and its ancestral population at some time $T$ in the past,  and suppose that we give ancestral individuals sub-population labels, defining \emph{source} populations. These labels are typically chosen to represent subgroups that have increased genetic homogeneity due to cultural or geographic reasons.  Then a simple summary of the demographic trajectory of a sampled allele is the source population from which it originated. We say that an individual is `admixed'  if it draws ancestry from multiple source populations--thus admixture is not an intrinsic property of individuals, but depends on our choice of labels and time $T$. An example of sub-population labels often used to study human populations in the Americas are the European, Native American, and West African populations prior to the advent of massive intercontinental travel.   Many routines have been proposed to infer the source population along the genome of admixed individuals \cite{Ungerer:1998p5534,Tang:2006p1488,Falush:2003p474,Hoggart:2004p3838,Patterson:2004p3833,Sankararaman:2008p4900,Bercovici:2009p5708,Price:2009p3896}. These typically proceed by locally matching an admixed genome to panel populations chosen as proxies for the source populations, revealing a mosaic of tracts of continuous ancestry (Figure \ref{karyogram}). In this work we use PCAdmix \cite{abra_s_thesis}, a heuristic approach for local ancestry inference. PCAdmix first divides the genome in windows of typical width of $10kb$ to $50kb$. For each window, the probability that the sample haplotype originates from any of the panel populations is estimated based on the position in PCA space. Finally, PCAdmix uses these probabilities as emission probabilities of a hidden Markov model and ancestry is inferred via Viterbi decoding.  

Local ancestry patterns have been used to identify disease loci [see  \cite{Seldin:2011p5719} and references therein] and to search for regions experiencing selection \cite{Tang:2007p3919,Bhatia:2011p4903}. They also provide hints about the history of migration \cite{Pool:2009p194}. The purpose of this article is to understand and model the observed ancestry patterns based on detailed demographic models, to learn about human demography, and to empower selection and association scans. In particular, we are interested in the length distribution of the continuous ancestry tracts, and the variation in ancestry proportions across chromosomes and individuals. 

A dominant stochastic process leading to these patterns is recombination, which tends to break down segments of continuous ancestry in admixed individuals. As a result, the length of continuous ancestry tracts tends to be shorter for more ancient admixture.  The tract length distribution is sensitive to details of recent migration (i.e., tens of generations), and is thus complementary to analysis based on the joint site-frequency spectrum \cite{Gutenkunst:2009p31,Gravel:2011p4622}, which is more sensitive at time scales of hundreds to thousands of generations. 

Recently, Pool and Nielsen \cite{Pool:2009p194} proposed a model in which a target population receives migrants from a source population, initially at a constant rate $m_2$. Starting at a time $T$ in the past, the rate changes to $m_1$.  In this model, back migrations are not allowed, recombinations within migrant chromosomes are neglected, and tracts shorter than a cutoff value are forgotten (since migration occurs  over an infinite period, this is necessary to avoid having a genome completely replaced by migrants). Assuming that recombinations occur according to a Poisson process, these approximations allow for an analytical solution for the distribution of tract lengths, which was used to infer demographic events in mice \cite{Pool:2009p194}. This model is limited to admixture proportions weak enough so that recombinations between migrant chromosomes are unlikely. A second limitation is that the model assumes two epochs of constant migration rate, which might or might not be the most appropriate for a given population. The special case $m_2=0$ has been used to infer demographic histories in humans for North African individuals \cite{Henn:2012p6117}. 

Here we propose a more general approach to predict the distribution of tract lengths that can accommodate both time-dependent and strong migration. This approach builds on that of Pool and Nielsen \cite{Pool:2009p194} but introduces multiple improvements. First, general time-dependent migrations can be considered. Second, recombinations between tracts of the same ancestry are not neglected, allowing for the modeling of strong migration and the simultaneous study of tracts of multiple ancestries. Third, chromosomal end effects are modeled explicitly. Fourth, our model can be modified to incorporate errors in tract assignments. \textcolor{black}{As in the Pool and Nielsen approach, we model recombination as a Poisson process with a unit rate per Morgan, and the recombination map is taken to be identical across populations [a reasonable approximation at the cM scale \cite{Wegmann:2011p6036}]. To perform demographic inference, we further require that local ancestry inference can be performed to high accuracy using one of the methods mentioned above. Whether this can be done depends on the degree of divergence of the ancestral populations (or sources), the availability of data for panel populations that are good proxies for the sources, and the possibility of accurately phasing diploid genomes.}

Admixture history also leaves a trace in the variance in admixture proportions across individuals, as stochastic mating and recombination tend to uniformize ancestry proportions with time \cite{Verdu:2011p5421}. Generalizing the models of \cite{Ewens:1995p5531,Verdu:2011p5421} to include the effects of recombination in a finite genome and drift,  we show that after a discrete admixture event, the variance decays in time in three consecutive regimes, first exponentially as differences in individual genealogies average out, then linearly as recombination creates shorter tracts, and finally exponentially again as drift fixes local ancestry along a chromosome. A simple approximate equation captures all three regimes accurately. By contrast, variance in continuous migration models is dominated by the first regime, and the expressions from the model of Verdu and Rosenberg \cite{Verdu:2011p5421} are reasonably accurate (see Appendix 3).


In general, distinguishing the effects of population structure and time-dependent patterns of gene flow is not straightforward, and the inference problem is prone to overfitting, as is the case, e.g., for inference based on the site frequency spectrum \cite{Myers:2008p2443}. However, our analysis shows that tract lengths, and more generally ancestry correlation patterns, can help resolve subtle differences in patterns of historical gene flow. An implementation of the proposed methods for tract length modeling, called \emph{tracts}, is available at http://tracts.googlecode.com.

\section{Theory}

\subsection{Admixture models: definitions and global properties}

\begin{figure}
\scalebox{.7}{\includegraphics{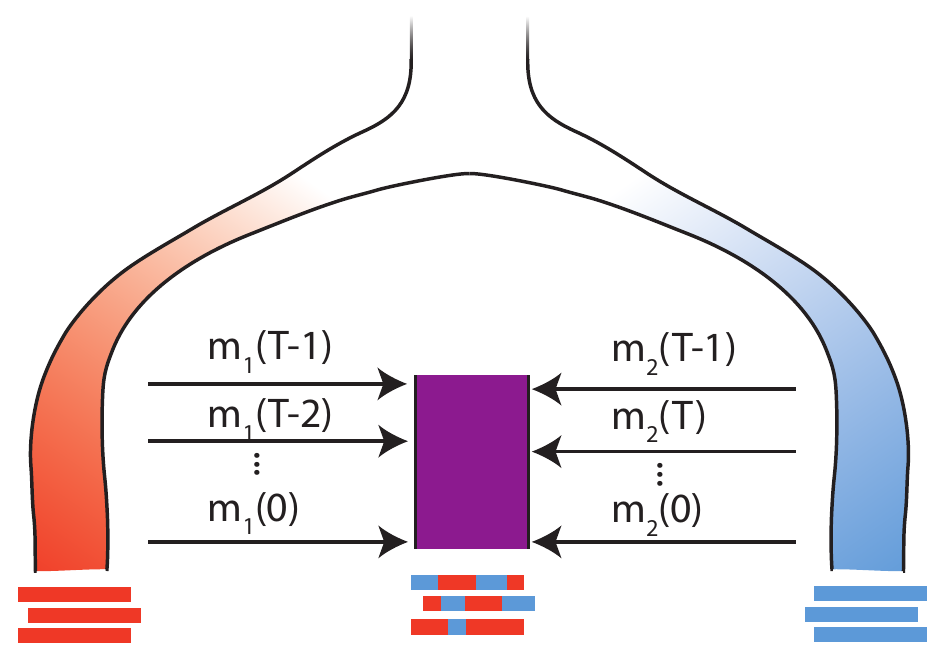}}
\hspace{1cm}
\scalebox{.7}{\includegraphics{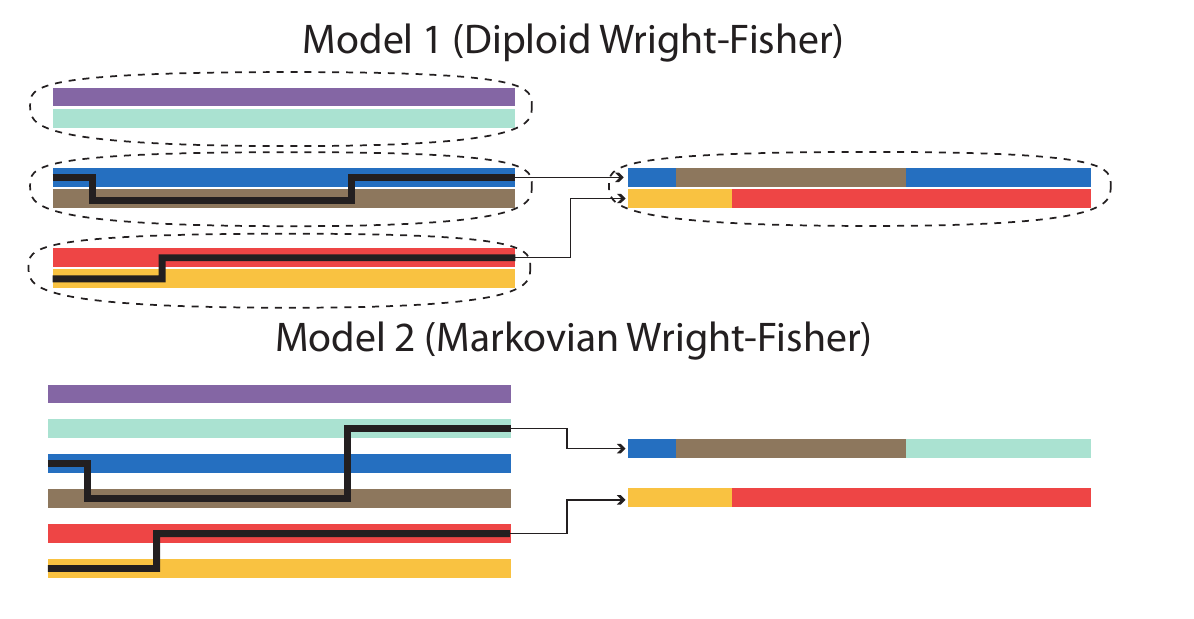}}

\caption{\label{greenhouse} 
(Left) Illustration of an admixture model starting at generation $T-1,$ where the admixed population (purple) receiving $m_i(t)$ migrants from diverged red ($i=1$) and blue ($i=2$) source populations at generation $t$. If these are statistically distinct enough, it is possible to infer the ancestry along the admixed chromosomes. Independent of our statistical power to infer this detailed local ancestry, the mosaic pattern may leave distinct traces in genome-wide statistics, such as global ancestry or linkage patterns. 
\textcolor{black}{(Right)  Gamete formation in two versions of the Wright-Fisher model with recombination. In Model 1, diploid individuals are generated by randomly selecting two parents, and generating gametes by following a Markov paths along the parental chromosomes. In Model 2, gametes are generated by following a Markovian path across the parental allele pool. Both models have the same distribution of crossover numbers, and are equivalent for genomic regions small enough that multiple crossovers are unlikely. Model 1 is more biologically realistic, and is used in the simulations, whereas Model 2 is more tractable, and is used for inference and analytic derivations.}}
\end{figure}

We wish to construct a model for the admixture of diploid individuals that takes into account recombination, drift, migration, and finite chromosome length. Since a full coalescent treatment of these effects is computationally prohibitive \cite{Griffiths:1996p4904}, we wish to simplify the model to consider only the demography of our samples up to the first migration event, $T$ generations ago. We label generations $s\in \{0,1,2,\ldots,T-1\}$, and the total fraction of the population $m(s)$ that is replaced by migrants in a generation $s$ can be subdivided in contributions $m_p(s)$ from $M$ migrant populations: $p\in\{1,\ldots,M\}$. We treat the replacement fraction $m_p(t)$ as deterministic, while the replaced individuals are selected at random (see Figure \ref{greenhouse}). 
Generations follow a Wright-Fisher model with random mating in a population with $2 N$ genome copies, each with $K$ finite chromosomes of Morgan length $\{L_i\}_{i=1\ldots,K}.$ We consider two different variations of the Wright-Fisher model with recombination.  

\textcolor{black}{
The first variation (Model 1) is meant to be the most biologically motivated and will be used for all simulations. Starting from a finite parental diploid population of size $N$, we first replace $m(s) N$ randomly selected individuals with diploid migrants. Diploid offspring are generated by drawing one gamete from each of two randomly selected diploid parents. Gamete formation is a Markov path with transition rate of one transition per Morgan across the two parental chromosomes (see Figure \ref{greenhouse}). }

\textcolor{black}{Model 1 results in long-range, non-Markovian correlations along the genome. This complicates the modeling without necessarily having a large effect on most global statistics. We will therefore also consider a more tractable model (Model 2) in which gametes are drawn from the migrant populations with probability $m(s)$, and are otherwise generated by following a Markov path along all non-migrant parental gametes (see Figure \ref{greenhouse}). The reason for singling out new migrants is that it is possible to generate their gametes as in the more realistic Model 1, without sacrificing tractability. Model 2 may not capture all long-range correlations in ancestry but it has the correct distribution of crossovers, and for small portions of the chromosomes is very similar to Model 1: the only difference is that each draw from the parental gamete pool is independent in Model 2, whereas the fact that a diploid individual can have multiple offspring induces a small degree of correlation between draws in Model 1. Unless otherwise stated, we calculate all population-wide statistics after the migration step, but before gamete generation. }

\textcolor{black}{Model 2 is reminiscent of the Li and Stephens copying model \cite{Li:2003p6153} used in HAPMIX \cite{Price:2009p3896}, as it also neglects back-and-forth recombinations due to multiple crossovers during a single meiosis. The purpose of the models are different, in that the current Markov models attempt to simulate gamete formation from parental chromosomes and represent evolution in time, whereas the Li and Stephens model attempts to simulate an unobserved haplotype based on haplotypes from the same generation.  The Markov ancestry transition model used in HAPMIX (and many other local ancestry inference software) corresponds to a special case of Model 2 when each population contributes migrants at a single generation.}

Local ancestry patterns are sensitive to the three stochastic processes of migration, recombination, and random genetic drift. Where possible, we take all three effects into account. By contrast, we do not model the effects of population structure, of selection, and of population size fluctuations. We derive our results under the assumption that local ancestries can be determined exactly; the effect of mis-identification are discussed throughout, together with possible correction strategies.

Given a history of migrations, it is relatively straightforward to calculate the expected population averages for ancestry proportions and tract lengths. If $m(s)$ is the total fraction of the population that is replaced by migrants, $s$ generations ago, with $m_i(s)$ from population $i$, \textcolor{black}{  the expected ancestry from population $i$ at a time $t$ in the past is the sum over generations $s$ of migrant contributions $m_i(s)$ weighted by the survival probability $\prod_{s'=t}^{s-1}\left(1-m\left(s'\right)\right)$  to time $t$. }  After the migration step, the ancestry proportions are:
$$\alpha_i(t)=\sum_{s=t}^\infty m_i(s) \prod_{s'=t}^{s-1}\left(1-m(s')\right).$$

We can follow a similar procedure to obtain the expected density $w_{ij}$ of ancestry switch-points from population $i$ to population $j$ per Morgan,  \textcolor{black}{replacing the amount of new migrants $m_i(s)$ by the density of new switch-points, which are proportional to the recombination rate (assumed constant with unit rate in genetic units) and the expected fraction of the genome $h_{ij}(s)$ that is heterozygous with respect to ancestries $i$ and $j$ after generation $s$.} In the gamete pool, we find:
$$\mathbb{E}\left[ w_{ij}(t)\right]= \sum_{s=t}^\infty h_{ij}(s) \prod_{s'=t}^{s-1}\left(1-m(s')\right).$$
The ancestry heterozygocity $h_{ij}$ can be evaluated using a recursive equation [such as Equation \eqref{recursion}], as in the case of allelic heterozygocity. In the absence of drift, $h_{ij}(s)=(1-m(s))  \alpha_{i}(s+1) \alpha_{j}(s+1)$.  In the population (before gamete generation), the sum over $s$ starts at $t+1$ rather than $t$. The expected number of switch-points per Morgan at time 0 is therefore
$$w_{ij}\equiv \mathbb{E}\left[ w_{ij}\left(0\right)\right]= \sum_{s=1}^\infty h_{ij}(s) \prod_{s'=0}^{s-1}\left(1-m(s')\right).$$

To estimate the expected tract length $\mathbb{E}\left[x_i(t=0)\right]$ for ancestry $i$ on a chromosome of length $L$, we divide the expected length covered by this ancestry, $\alpha_i(0)*L$,   by the expected number of tracts of this ancestry, which is $\frac{L}{2} \sum_j  w_{ij}   +\alpha_i(0)$ since each tract must begin and end by an ancestry switch or by the end of the chromosome. We thus find: 
%
%
$$\mathbb{E}\left[x_i(t=0)\right]=\frac{2\alpha_i(0)L}{L \sum_j w_{ij}  +2 \alpha_i(0)}.$$

If the demographic model under consideration has a single parameter, such as the timing of a single pulse of migration, demographic inference can proceed from this single estimate. However, the mean tract length may be largely dependent on the number of very short tracts which are difficult to detect; this statistic is therefore sensitive to false-positive and false-negative ancestry switches. Here we are interested in studying more detailed models of migration and their impact on tract length distribution.

 \subsection{Tract length distribution}

\begin{figure}
\scalebox{1}{ \includegraphics{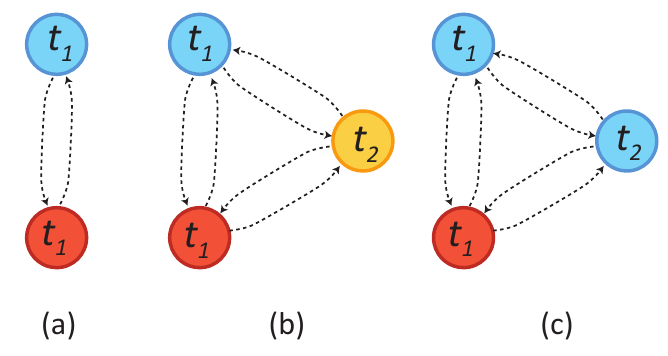}}
\caption{\label{2state}\textcolor{black}{(a) A two state Markov Model for ancestry along a chromosome for a single pulse of migration at time $t_1$. Tract lengths distributions are exponential. (b) A three-population Markov Model with a pulse of blue and red ancestry at time $t_1$ followed by a pulse of migration from the yellow population at time $t_2$. All tract length distributions are exponential. (c) A two population model in which the blue population contributes migrants at generation $t_1$ and $t_2.$ The distribution of blue ancestry tracts is no longer exponential, as we cannot detect transitions between blue states. }}
\end{figure}

For illustration, we first consider a source population (Blue), and a target population (Red), with a single, infinitely long diploid chromosome. At generation $t =T-1$, a fraction $m$ of population Red is replaced by individuals from population Blue. Consider the Markovian Wright-Fisher Model discussed above (Model 2). \textcolor{black}{In this model, the position of the closest recombination to either side of a point along an infinite chromosome is exponentially distributed and there is no memory of previously visited states along a chromosome. The chromosomes resulting from this admixture process can therefore be modeled as a continuous-time Markov Model with a Red and a Blue state (Figure \ref{2state}(a)), where each recombination event corresponds to a Markov transition and the continuous Markov `time' corresponds to the position along the chromosome. }
The transition rate out of a state in this model is proportional to the number of recombinations, namely $t-1$ per Morgan: since recombinations within first-generation migrants do not induce ancestry changes, and we suppose that we sequence somatic cells at generation $0$, there can only be recombination during gamete formation at generations $1,...,t-1.$ 
If a recombination occurs, the probability of transitioning is $m$ to the Red state, and is $(1-m)$ to the Blue state. 
 
We are interested in the length distribution of continuous segments in the Blue or Red ancestry, independent of the number of within-ancestry transitions, which are difficult to detect. We avoid these complications by setting the self-transition rates to zero: this does not affect the trajectories, but now all transitions change the ancestry. We therefore have the model shown in Figure \ref{2state} (a), and the distribution of tract lengths $\phi_i(x)$ is equal to the exponential distribution of distance between Markov transitions: 
\begin{equation}
\begin{split}
\phi_R(x)&=m (t-1) e^{-m (t-1) x}\\
\phi_B(x)&=(1-m) (t-1) e^{-(1-m) (t-1) x}.
\end{split}
\end{equation}
Note that the distribution is ill-defined for $t=1$, since this situation produces tracts that are infinite in the infinite-chromosome limit. 

\subsubsection{Multiple populations, discrete migration}

As long as the migration from each population is limited to a single generation and the target population is infinitely large, Model 2 produces Markovian trajectories along ancestry states. To see this, consider a point $x$ along the genome in a segment from ancestry $p$ that arrived $t$ generations ago. As before, the distance to the first recombination event downstream from $x$ is exponentially distributed (with rate $t-1$), and the timing $\tau$ of the recombination is uniform on $(1,t-1)$. Moreover, since gametes in Model 2 are formed by following a Markov path in the parental gamete pool, the probability of observing ancestry $p'$ downstream from the recombination is proportional to the ancestry proportions in the parental pool at the time $\tau$ of the recombination.  Thus we have the discrete transition rate $$M(p\rightarrow p')=\sum_{\tau=1}^{t-1} P(p'|\tau)P(\tau|p)=\sum_{\tau=1}^{t-1} \frac{\alpha_{p'}(\tau+1)}{t-1},$$ which depends only on the time of arrival $t$ of ancestry $p$. We note that the Markov property over ancestry states would be lost in Model 1, because the state downstream of the recombination is correlated with upstream states. Drift reduces the transition rates and also breaks the Markov property: mitigation strategies are discussed in Appendix 1.  The Markov property over ancestry states is also lost if a population contributes migrants over many generations, and our next step is to restore the Markov property in this situation by extending the state space. 

\subsubsection{General incoming migration in the absence of drift} 
We now allow for general incoming migration histories that start at a time $T-1$ in the past. For each generation $t\in\{0,\ldots,T-1\},$ a fraction $m_p(t)$ of the individuals from the target populations are replaced by individuals from the source population $p$, with $m(t)=\sum_p m_p(t)\leq 1$. We further impose that the first generation is composed of non-admixed individuals: $m(T-1)=1$. Since the ancestry switches are no longer Markovian in the general migration case, it is convenient to consider states defined by both ancestry $p$ and time of arrival $t$. Intuitively, we may imagine that we have a large number of migrant populations $(p,t)$, each contributing migrants over a single generation (see Figure \ref{2state} (b) and (c)). Here the Markov property is maintained, but ancestry states can now correspond to multiple Markov states.
%
%
%



We first calculate the transition rates between states $(p,t)$ as we did for the discrete migration case. First, the probability of encountering state $(p,t)$ downstream from a recombination that occurred at time $\tau$ is
$$P(p,t|\tau)=\Theta(t-(\tau+1)) m_p(t) \prod_{t'=\tau+1}^{t-1} \left(1-m(t')\right), $$
where  $$\Theta(s)=\begin{cases}1 &s\geq0\\0 &\mbox{otherwise}   \end{cases} $$ 
is the Heaviside function.

As before, given a point $x$ in state $(p,t)$, the position of the next downstream recombination is exponentially distributed with rate $t-1$, and the time of this recombination is uniformally distributed on $(1,t-1)$. In the two Wright-Fisher models considered here, states on either side of the recombination are uncorrelated, and we can write the discrete transition probabilities
$$R(p,t \rightarrow p',t')=\sum_{\tau=1}^{\min(t,t')-1} \frac{P(p',t'|\tau)}{(t-1)}, $$
which is independent of $p$.
The continuous transition rate is obtained by multiplying the discrete transition rate by the continuous overall transition rate $t-1$:
\begin{equation}
\label{dlrate}
Q(p, t \rightarrow p',t')=m_{p'}(t')\sum_{\tau=1}^{\min(t,t')-1}  \prod_{s=\tau+1}^{t'-1} \left(1-m(s)\right). 
\end{equation}
These transition probabilities are valid for both Wright-Fisher models in the infinite-population size limit. Since Model 2 is Markovian, these transition rates are sufficient to fully specify the ancestry state model. 

Given the transition matrix $Q$, we can use standard tools for the study of Markov chains to efficiently estimate the length distribution of excursions on Markov states corresponding to a single ancestry. In Appendix 2, we first derive results under the approximation that chromosomes are infinitely long. We account for finite chromosomes by studying the distribution of tract lengths in finite windows, randomly chosen along the infinite chromosomes. We thus obtain a distribution of tracts $\phi_p(x)$ for each population $p$. To compare these predictions to observed data, a computationally efficient strategy is to bin data by tract length, and treat the observed counts in each bin as an independent Poisson variable with mean obtained by integrating $\phi_p(x)$ over the bin range.

Short ancestry tracts are likely to have both elevated false positive and false negative rates, and inference based on such tracts is likely to be biased, whereas longer tracts can be detected with increased confidence. Following \cite{Pool:2009p194}, we therefore perform inferences using only tracts longer than a cutoff value $C$. We should emphasize that a large number of uniformly distributed spurious short tracts may still impact the distribution of longer tracts, making non-exponential distributions look more exponential. Therefore, significant assignment error may cause an underestimation of the amount of continuous migration. By contrast, drift would tend to reduce the transition rates and cause underestimates of the time since admixture (see Appendix 1).

\subsection{Variance among individuals}

We now consider the variance among individuals in total migrant ancestry $X^p$ from population $p$, measured as a proportion of the Morgan length of the genome whose origin is from $p$. The variance in ancestry can be separated in two components, which we label the genealogy variance and assortment variance. The genealogy variance is due to a different number of migrant ancestors; if a randomly chosen fraction $m$ of the population is replaced by migrants at each generation, a fraction $m^2$ of individuals will have two migrant parents, $2m (1-m)$ will have one migrant parent, and $(1-m)^2$ will have none. The assortment variance accounts for the fact that two individuals with the same genealogy can vary in their genetic ancestry proportions, since not all ancestors contribute the same amount of genetic material to an individual. Recombination and the independent  assortment of chromosomes tend to reduce such variance. 

We can use the law of total variance, conditioning over the genealogies $g$, to isolate these two contributions to the variance $\var(X^p):$
$$\var(X^p)=\var_g\left[\mathbb{E}(X^p|g)\right]+\mathbb{E}_g\left[\var(X^p|g)\right].$$
Here $\mathbb{E}\left[X^p|g\right]$ is the fraction of migrant ancestry from population p, based on the genealogy $g$. Alternatively, this can be thought of as the infinite-sites expectation for the ancestry proportions. The first term therefore represents the genealogy variance in ancestry, whereas the second term represents the assortment variance. Because of random chromosome assortment, the variance in ancestry between chromosomes is informative of the assortment variance. We discuss in Appendix 3 how, in the absence of drift, the variance can thus be broken down in these two components without requiring a demographic model. We discuss below how to obtain expectations for each components given a specific demographic model.

\subsubsection{Genealogy variance}

 To ease calculations of the genealogy variance, we neglect correlations due to overlap between individual genealogies, and describe each individual as being sampled from an independent genealogy (in a randomly mating population, this amounts to neglecting drift). In this model, the genealogy variance $\var_g\left(\mathbb{E}\left[X^p|g\right]\right)$ is easily calculated. Considering the genealogy $g$ of a non-migrant sample up to $T$  generations ago (we label the current generation $0$, and the generation with the first migrants $T-1$), we first note that 
 $$\mathbb{E}\left[X^p|g\right]=\frac{1}{2^{T-1}}\sum_{i=1}^{2^{T-1}} z^p_i,$$
  where $z^p_i$ is 1 if there has been a migrant on the lineage leading from the root to leaf $i$, $0$ otherwise. Results with continuous admixture since time immemorial can be obtained by taking a limit $T\rightarrow \infty$. In such cases, the approximation of independent pedigrees eventually breaks down, but the resulting expression might remain approximately correct if the majority of present day genomes originate from recent migrants.
 
The expectation  over genealogies $g$ and assortments $\mathbb{E}_g\left[\mathbb{E}\left[X|g\right]\right]$ is then $\alpha_p(0)$. The calculation of $\mathbb{E}_g\left[\mathbb{E}\left[X|g\right]^2\right]$ is also straightforward if we can calculate \textcolor{black}{the expectation $\mathbb{E}_g\left[z^p_i z^p_j\right]$. For $z^p_i z^p_j$ to be nonzero, we must have had a migrant either on the common branch leading to the two leafs $i$ and $j$, or one migrant on each of the separate branches: }
%
\begin{equation}
\begin{split}
 \mathbb{E}_g\left[z^p_i z^p_j\right]=\sum_{s=0}^{T-1-d_{ij}} m_p(s) \prod_{s'=0}^{s-1} (1-m(s'))+\alpha_p^2(T-d_{ij}) \prod_{s=0}^{T-1-d_{ij}} (1-m(s)) \equiv  e(d_{ij}),
\end{split}
\end{equation} 
 with $d_{ij}$ is half the tree distance between leafs $i$ and $j$. \textcolor{black}{Then we can write the sum over half-distances, weighted by the number of leaf pairs at each distance}:  
 \begin{equation}
 \mathbb{E}_g\left[\mathbb{E}\left[X|g\right]^2\right]=\sum_{d=1}^{T-1} 2^{d-T} e(d)+\alpha(0)/2^{T-1}.
 \end{equation}
 Since $ \mathbb{E}_g\left[\mathbb{E}\left[X|g\right]\right]=\alpha(0),$ we have
 $$\var_g(\mathbb{E}[X|g])=\sum_{d=1}^{T-1} 2^{d-T} e(d)+\alpha(0)(\frac{1}{2^{T-1}}-\alpha(0)).$$
  In the two-population pulse model, with $m_{p=1}(t)=m\delta_{t,T-1},$ we have the expected  $\var_g(\mathbb{E}[X|g])=\frac{m (1-m)}{2^{T-1}},$ with a rapid exponential decay of the variance as a function of $T$. By contrast, if we have continuous migration of population $p$ in a target population, with, $m^p_i=m\Theta(T-i-1),$ the variance reads
\begin{equation}
\begin{split}
\var_g(\mathbb{E}[X|g])=\frac{2^{-(T-1)} m (1-m)^{T} \left([2(1-m)]^{T}-1\right)}{1-2 m},
\end{split}
\end{equation}
with a more complex dependence of the variance on $T$. Finally, in the case where two populations provide respectively $p m$ and $(1-p)m$ migrants to a target population at each generation since the beginning of time, we have the simple expression:
 \begin{equation}
\var_g(\mathbb{E}[X|g])= \frac{2 p (1-p) m}{1+m}.
\end{equation}
This expression supposes that the variance is calculated after migration occurs. If variance is calculated before replacement by migrants, the factor of two disappears, and we recover equation (47)  in \cite{Verdu:2011p5421}.


\subsubsection{Assortment variance}

To study the global ancestry variance due to assortment, a natural starting point is to consider the ancestry variance at a particular point in the genome. In a randomly mating population with two ancestries, the variance in ancestry at a site is $h/2$ , where $h$ is the ancestry heterozygocity at that site. The ancestry heterozygocity can be calculated using the same recursive strategy commonly used for allelic heterozygocity [equation \eqref{recursion}]. 
The case of three or more ancestries can be reduced to two ancestries by singling out one ancestry and pooling the others.  As a specific example, in the case of a pulse migration with migration rates $m$ and $1-m$ at generation $T-1$, the heterozygocity at generation zero is
\begin{equation}\label{hettime}
h_0=(1-\frac{1}{2N})^{T-1} 2 m (1-m).
\end{equation}

We wish to combine these local variances into an expression for the genome-wide variance. \textcolor{black}{In Appendix 3 we provide a derivation of the expected ancestry variance using Markov Models. Here, to obtain a simple approximation for the migration pulse model, we imagine that the length of the genome is divided in $n$ tracts by uniformly drawing $n-1$ separators. We suppose that the ancestry is chosen independently on each segment, with variance $h_0/2$. Then the variance in ancestry in the large-$n$ limit is  } 

$\mathbb{E}_g\left[\var(X^p|g)\right]\simeq \frac{h_0}{n}.$

\textcolor{black}{The effect of drift is therefore captured by the decay of ancestry heterozygocity with time, whereas the effect of recombination is captured by the number of independent tracts $n$,}  which is proportional to the number of recombinations. In the case of a pulse of migration $T$ generations ago without drift, we write $n=1+ (T-2) L_i$ for a single haploid chromosome (the $1$ accounts for the chromosome edge, and can be neglected for large $T L_i$), and $2K +2 (T-2)L$ for a diploid genome with $K$ chromosome pairs of total length $L=\sum_i L_i$. Thus the total variance reads: 
\begin{equation}
\var(X^p)=\frac{m(1-m)}{2^{T-1}}+\frac{2m (1-m) (1-1/2N)^{T-1} }{2K +2 (T-2)L}.
\label{varpulse}
\end{equation}
Even though it neglects the effect of drift on the number of independent tracts $n$, this expression provides excellent quantitative agreement with simulations over multiple regimes (Figure \ref
{variances}). If we model the variation over time of the population ancestry proportion as a random walk with decreasing step size $\frac{\var(X^p)}{N},$ the variation will be dominated by the genealogy variance, which after an infinite time contributes a finite variance of $\sigma^2=\frac{m (1-m)}{N}$. 
Thus for an initial population of 100 individuals divided equally between two ancestries, we can expect the final ancestry proportions to be $0.5 \pm 2 \sigma = 0.5 \pm 0.1$, a relatively modest uncertainty given the small population size.  Assortment variance for continuous migration models is discussed in Appendix  3.




 
 \section{Comparison with simulation and experimental data}

In this section we first present results of Wright-Fisher simulations, comparing our model predictions to the simulation results. We then consider the HapMap African-American panel, for which we performed local ancestry inference and analyzed the tract length distribution.
\subsection{Tract lengths}

\begin{figure}
 \scalebox{.75}{\includegraphics{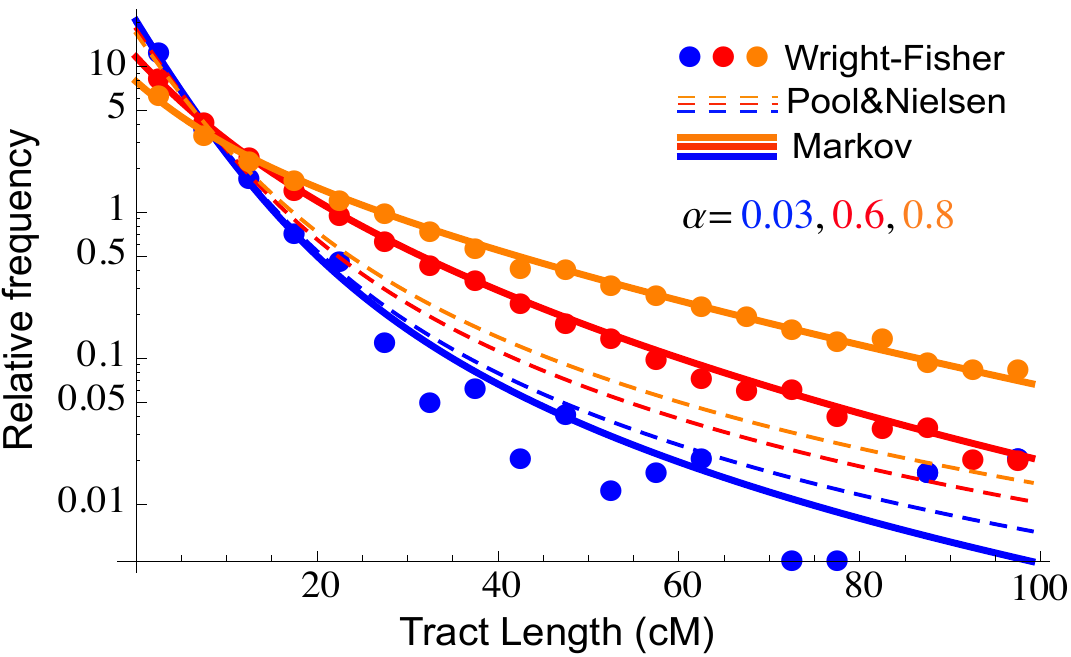}}
 \caption[figPN]{\label{figPN} \textcolor{black}{Comparison of the Markov Model, the Pool and Nielsen prediction\protect{\cite{Pool:2009p194}}, and Wright-Fisher simulation for migrant tract length distributions. Each dot represents the normalized number of ancestry tracts whose length is contained in one of 20 bins.  The simulation followed $10000$ chromosomes over 30 generations, with constant migration rates $m=0.001, 0.03, 0.05$ giving rise to final ancestry fractions of $\alpha=0.03,0.6,0.8$. Since recombination between migrant tracts were neglected in \protect{\cite{Pool:2009p194}}, the results depart significantly from simulation at high migration, whereas the Markov Model is accurate in the three regimes.}}
 \end{figure}

\textcolor{black}{We performed a $30$-generation diploid Wright-Fisher simulation (using Model 1, see Figure \ref{greenhouse}) of $10000$ chromosomes of length 1 Morgan with continuous gene flow from population 1 into a population initially composed of individuals from population 2. We considered three different migration intensities, namely $m_1=0.001, 0.03$, and $0.05$ per generation. We kept track of the ancestry of each segment during the simulation, so that the continuous ancestry tracts could easily be tabulated. On Figure \ref{figPN}, we compare the observed histograms of tract lengths for population 1 (dots) to predictions from equation 10 in \cite{Pool:2009p194} (dashed lines) and to predictions from the Markovian Wright-Fisher model (Model 2 on Figure \ref{greenhouse}), using rates from equation \eqref{dlrate} and implemented as described in Appendix 2 to account for finite chromosome length (solid lines). As expected, the predictions of the two models  are similar when migration rates are low, and differ substantially when we depart from the weak migration assumptions of the Pool and Nielsen model (see Figure \ref{figPN}). The Markov model predictions are in good agreement with the simulations over the range of models considered, including when the migrant population becomes the majority population. }

We now consider the HapMap African-American panel (ASW) \cite{InternationalHapMap3Consortium:2010p2210}, and focus on 20 unrelated samples that were trio-phased, to reduce biases due to phasing errors. We obtained local ancestry inferences using PCAdmix \cite{abra_s_thesis}, using 132 unrelated HapMap samples from Europe (CEU) and 204 from West Africa (YRI) as reference panels. We used windows of size $0.3$cM for the HMM 
and based our inferences on the number of tracts longer than $10$cM. We pooled tracts in 50 bins according to tract length (chromosomes with no ancestry switches were in a separate bin independent of the chromosome length), and calculated model likelihood assuming that counts in each bin are Poisson distributed with mean given by the model predictions for this bin.

\begin{figure}
\vspace{-1cm}
\scalebox{0.3}{\includegraphics{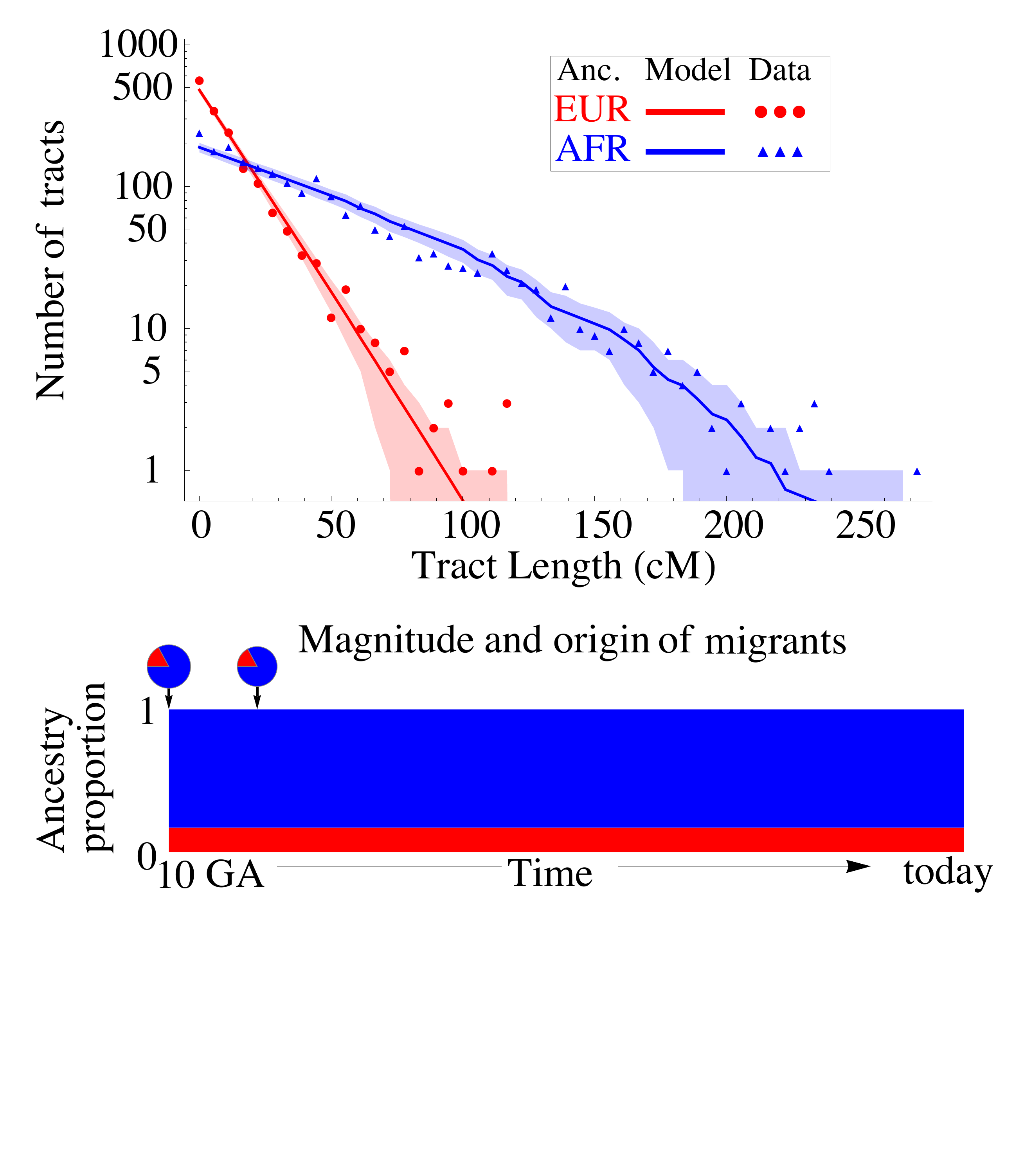}}
\vspace{-1cm}
\scalebox{0.3}{\includegraphics{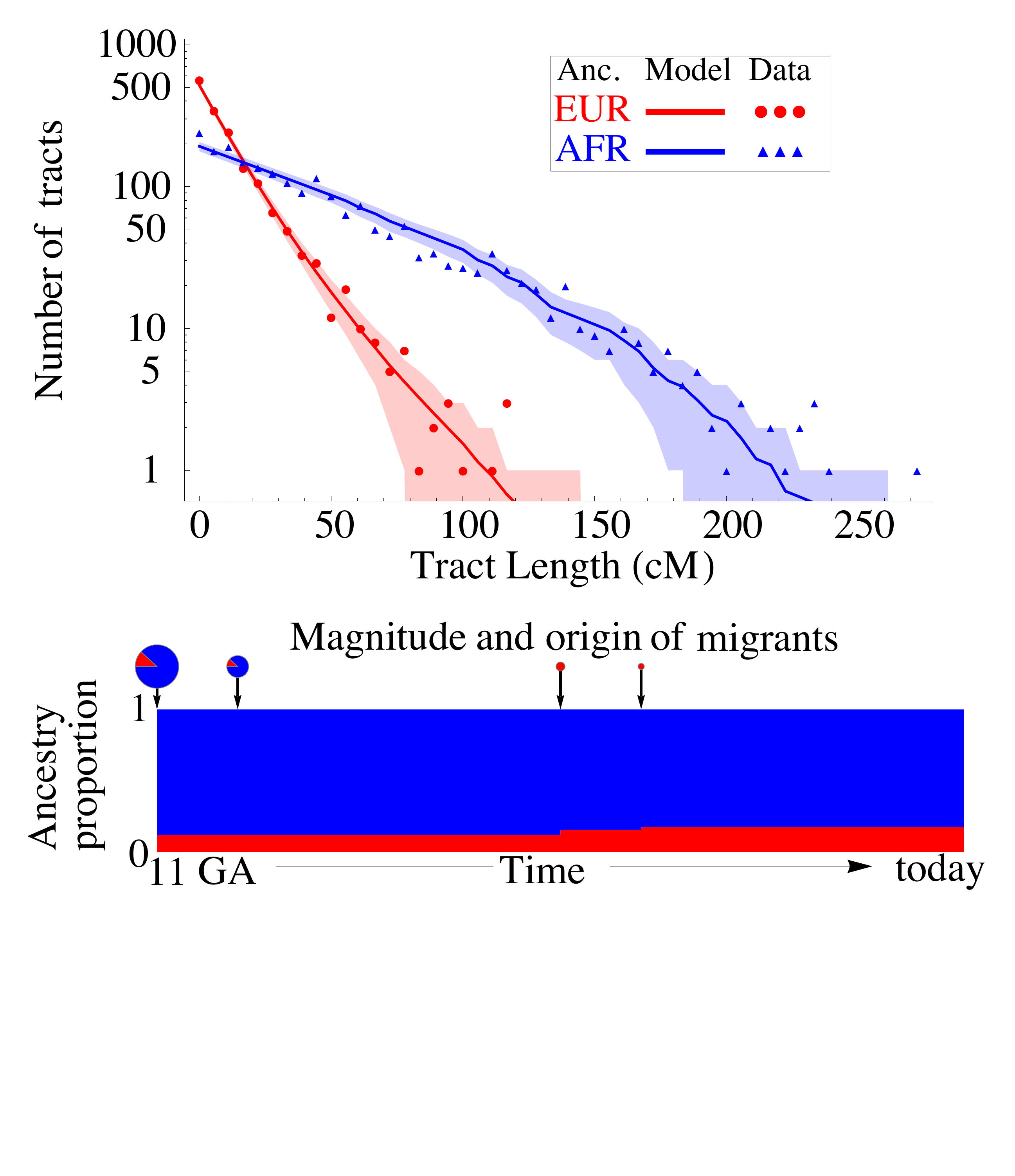}}
\vspace{-1cm}
\caption[tractInf]{\label{tractInf} Distribution of continuous ancestry tract lengths in $20$ HapMap African-American (ASW) trio individuals (as inferred by PCAdmix \protect{\cite{abra_s_thesis}}, a local ancestry inference software), compared with predictions from a single pulse migration model (Top) and a model with subsequent European migration (Bottom). Each dot represents the number of continuous ancestry tracts whose length is contained in one of 50 bins. The shaded area marks the $68.3\%$ confidence interval based on the model. The second model, in which over $30\%$ of European origin in the ASW samples is quite recent, provides a sufficiently better fit to justify the extra parameters (likelihood-ratio test, $p=0.002$). }
\end{figure}

We compared inferences based on 2 different models; (a) a `pulse' model, with a single migration event, and (b) a 2-pulse model, with a subsequent migration of Europeans (Figure  \ref{tractInf}). Model (b) has two additional parameters, corresponding to time and proportion of the subsequent European migration. A likelihood ratio test shows that $\ln (\mathcal{L}_b/\mathcal{L}_a) \simeq 7$. To establish the significance of the extra two parameters, we simulated 1000 random tract length distributions from the maximum likelihood model (a), and obtained maximum likelihood estimates for both models. The probability of obtaining such a likelihood ratio under model (a) is $p=0.002$.  

\subsection{Ancestry proportions and variance}

Simulations of 80 individuals, each with 22 autosomal chromosomes of realistic lengths (namely  2.78, 2.63, 2.24, 2.13, 2.04,
1.93, 1.87, 1.70, 1.68, 1.79, 
1.59, 1.73, 1.27, 1.16, 1.26, 
1.35, 1.30, 1.19, 1.08, 1.08, 
0.62, 0.73 Morgans, for chromosomes 1 to 22, respectively) and $30\%$ of initial admixture proportion, illustrate many of the effects predicted in variance models. The global ancestry proportions and fraction of sites heterozygous for ancestry fluctuate considerably over the first few generations, but the fluctuations decrease in time as ancestry proportions approach a fixed value and ancestry heterozygocity decays following equation \eqref{hettime}.



\begin{figure}
  \scalebox{.25}{\includegraphics{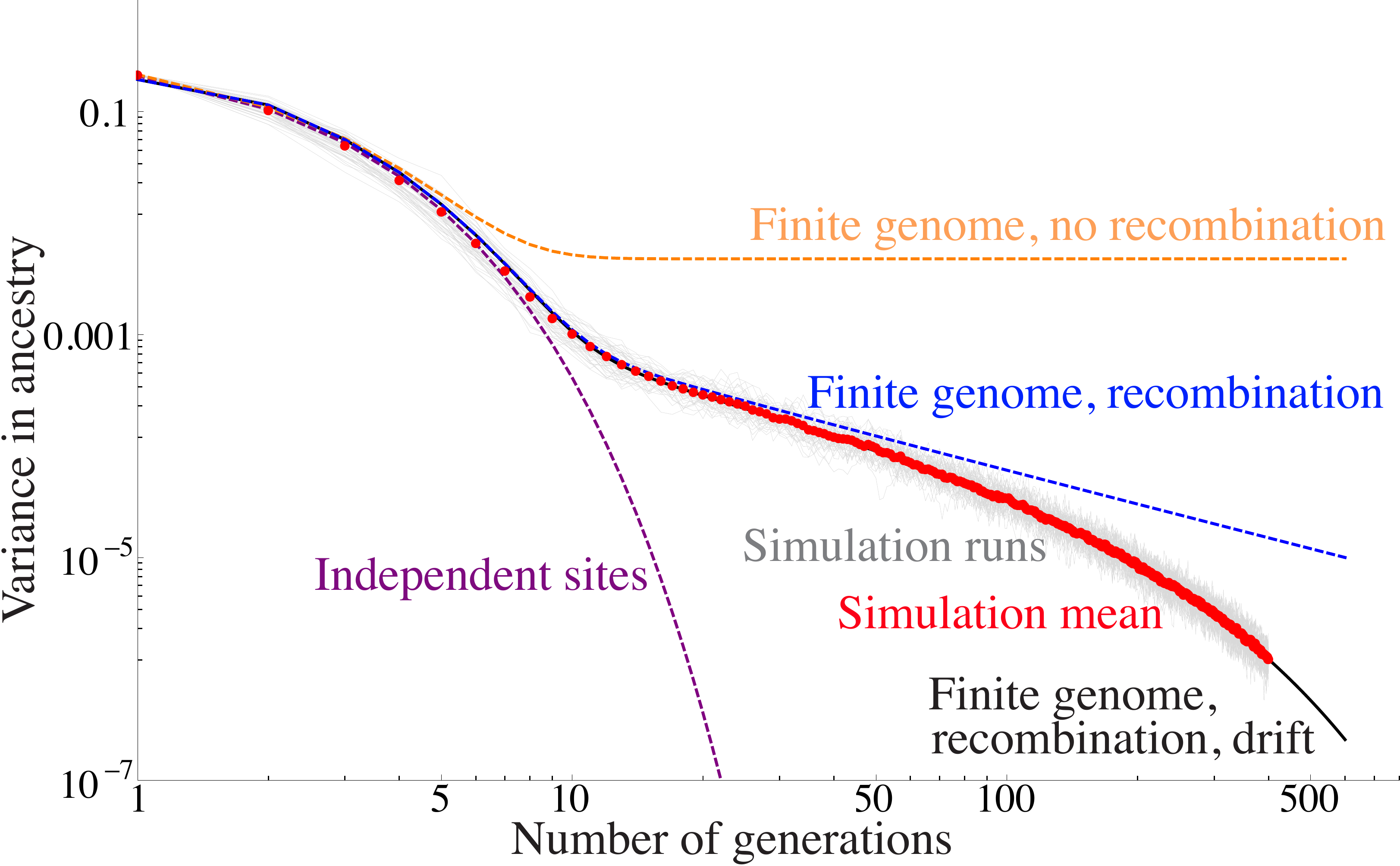}}
 \caption{\label{variances} \textcolor{black}{Comparison of 50 independent  Wright-Fisher simulations of a population of 80 samples and $30\%$ admixture proportion to predictions from increasingly detailed models.  We show the variance in ancestry across individuals for each simulations in pale gray, and the average over the simulations is shown as red dots. These are compared to predictions for an independent sites model (purple)  for a finite genome with 22 non-recombining chromosomes (orange), for a model with recombination (blue), and finally for a model with recombination and drift given by Equation \eqref{varpulse} (black). The latter model captures the variance in quantitative detail over three qualitative regimes}.} 
 \end{figure}

Figure \ref{variances}
shows that the variance in ancestry across individuals follows three different regimes; first, the variance is dominated by the genealogy variance, with a rapid exponential decay. After about 10 generations, the assortment variance starts to dominate, and decays polynomially due to recombination until drift becomes important, where an exponential decay is resumed, although at a much reduced rate. 
 
Equation \eqref{varpulse} captures these three regimes in quantitative detail--the average variance over 50 independent simulations follows the model prediction closely. The continuous migration case, where genealogy variance tends to dominate, is discussed in Appendix 3.

Comparing the ancestry variance from the African-American data to those predicted by the demographic models, we find that the pulse model predicts a genealogy variance of $0.0005,$ whereas the variance in the model with two distinct pulses is $0.002$. The total variance in the African-American sample is $0.0047$, of which we infer that $0.0041$ is due to genealogy variance (using the method described in Appendix 3). Thus the model with two pulses of migration is again more realistic than the single pulse model; the fact that it still underestimates the variance can be due to a combination of factors that have not been modeled: our demographic model may be underestimating low level, very recent migration because of the parameterization as two discrete pulses of migration; and both population structure and errors in ancestry assignment  may be adding to the observed variance.

 \section{Discussion}

\subsection{Limitations and possible improvements}

A limitation of all demographic inference methods is that the model space is very large, and the information available to learn about the models is limited. Thus we need to coarsely  parameterize model space at the risk of introducing biases. This is similar to the modeling of allele frequency distributions: even though the vast majority of scenarios are inconsistent with the data, the number of models that are consistent with the data remains large, and model-fitting often requires imposing additional simplifying assumptions. When applied to HapMap trio-phased African-American data (ASW), inferred parameters were reasonable and we found evidence for migration patterns that depart from the migration pulse with subsequent random mating which is at the heart of many approaches.  However, distinguishing between continuous migration and nonrandom mating remains challenging.


The demographic inference strategy we presented requires accurate local ancestry assignments. Since longer tracts contain more ancestry information, we expect the most significant types of mis-assignment to be short, spurious ancestry tracts, and the failure to identify real, short ancestry tracts. In the HapMap ASW data used here, the source populations are diverged enough that assignments are relatively reliable down to relatively short tracts, and indeed we find that the number of tracts predicted by the model is in good agreement with the data for the shortest tracts, even though these were not used in the fitting procedure. If the expected number of mis-identified short tracts is large enough that it will strongly affect the distribution of longer tracts (by introducing spurious breaks in longer tracts), the Markov Models should be modified to include mis-identification states, and transition rates could be estimated via simulated admixed individuals.



Alternatively, it is possible to circumvent the local ancestry inference step altogether by focusing on a derived statistic, such as the decay of correlation in ancestry informative markers with genetic distance. Such a method was proposed in \cite{Reich:2009p478}, for the case of pairwise ancestry correlations in a pulse migration model. Even though such an approach avoids possible biases due to local ancestry assignment, pairwise ancestry correlations become noisy as distance is increased, and are thus less sensitive to continuous gene flow patterns. The Markov models presented here provide a natural framework to generalize linkage-based models for more general admixture scenarios, as arbitrary order linkage statistics can be derived in the Markov framework. Furthermore, HMM approaches could be developed to model the complete mosaic pattern without the need to focus on summary statistics such as the tract length distributions and ancestry variances. Even though such approaches would be more computationally intensive, they may increase the accuracy of the inference, especially when assignment errors are important. 
\subsection{Conclusion}  
Overall, we found that the proposed models accurately describe the distribution of ancestry tract lengths and variances when compared to Wright-Fisher simulations. The models we used allow for general migration histories, yet are tractable and can be used for inferring demographic parameters in real data. They are therefore useful to improve our understanding of the consequences of gene flow and our ability to infer demography in populations with complex histories. Such populations have often been underrepresented in medical genetic  studies, in part because of complications in the modeling of genetic heterogeneity. As medical genetics sampling efforts strive to reduce this disparity, detailed models for genetic diversity will be increasingly important to make the most out of the resulting data.


\section{Acknowledgements}
I thank Carlos D. Bustamante, Jake K. Byrnes, Brenna M. Henn, Jeffrey M. Kidd and Damien Simon for useful discussions. This publication was made possible by NIH/NIGMS grant number 1 R01 GM090087-01 and NIH/NHGRI grant number U01HG005715.
\bibliographystyle{mychicago}
\bibliography{./tractlengthbib2,./tractlengthbib_byhand}
\pagebreak

\appendix

\section{Appendix 1:The effect of drift on ancestry transitions}
\label{drift}

Drift increases the probability that recombinations occur between segments of the same ancestry. In the infinite-time limit, ancestry will have fixed at every site, no more ancestry switches are created, and the tract length distribution is constant in time. In the presence of drift, the ancestry switches are no longer Markovian; if a recombination occurs between two IBD segments, it increases the posterior probability that the next recombination will also be between IBD segments. However, it is likely that a Markovian approximation will remain accurate for moderate drift if we take into account the reduced probability of ancestry-switching recombinations.

We first wish to obtain the fraction of recombinations that occur within segments  $(p,t),$ of ancestry $p$ having migrated at generation $t$, as these recombinations do not induce ancestry switches and will be most affected by drift. In other words, we want to find the fraction of sites that are homozygous for the ancestry $(p,t)$, and contrast this to the case with no drift. For this purpose, we consider all other ancestries as  a single allele, and in the first step we compute the total homozygocity of non-migrants in this system $s$ generations ago: $f^s_{p,t}$. \textcolor{black}{We write the usual recursive relation over generations, noting that a homozygous state in a Wright-Fisher model can be obtained in one of four parental situations: drawing the same non-migrant parent twice, drawing two non-migrant parents with the same ancestry, drawing one last-generation migrant and a non-migrant with the same ancestry, and finally drawing two last-generation migrants:}
\begin{equation}
\begin{split}
\label{recursion}
f^s_{p,t}=&\left(\frac{1}{ 2 N (1-m(s+1))} +\left(1-\frac{1}{2 N (1-m(s+1))}\right) f^{s+1}_{p,t}\right)\left(1-m(s+1)\right)^2 \\&+ m(s+1)\left(1-m\left(s+1\right)\right) \left(1-\alpha_{p,t}(s+2)\right)+m^2(s+1).
\end{split}
\end{equation}
This recursion can be initiated with the homozygocity one generation after $t$, namely  $f^{t-1}_{p,t}=m_p(t)^2+\left(1-m_p\left(t\right)^2\right).$ Finally, to get the fraction $c_{p,t}$ of nonmigrant sites that are homozygous for the $p,t$ ancestry at generation $s$, we write
\begin{equation}
2 \alpha_{p,t}(s+1)=2c_{p,t}(s)+1-f^s_{p,t}
\end{equation}
and solve for $c_{p,t}$:
\begin{equation}
\label{apt}
 c_{p,t}(s)=\frac{f^s_{p,t}-1}{2}+\alpha_{p,t}(s+1), 
\end{equation}
which reduces to $\alpha^2_{p,t}(s+1)$ in the drift-less limit.

In the drift-less case, the probability of the state to the right of a recombination depended only on the time of the recombination. Due to the possibility of recombining within segments identical by descent, this is no longer the case when drift is present. However, consider a given point $x$ in state $(p,t)$ along the genome. The distribution of the distance to the first recombination encountered upstream (or downstream) from $x$ is unaffected by drift. Thus the relationship between transition rates $Q$ and discrete transition probabilities $R$ is maintained: $Q(t,p\rightarrow t',p')=(t-1) R(t,p\rightarrow t',p')$ for $(t,p)\neq (t',p').$
If we indicate the state to the left or right of a recombination by a left- and right-pointing arrow, respectively, we write
\begin{equation}
\begin{split}
R(t,p\rightarrow t',p')&\equiv  P( (t',p')_{\rightarrow}|(t, p)_{\leftarrow})\\
&=\sum_{\tau=1}^{\min{(t,t')}-1} \frac{P( (t',p')_{\rightarrow}|\tau,(t, p)_{\leftarrow})}{t-1}\\
&=\sum_{\tau=1}^{\min{(t,t')}-1} \frac{P((t, p)_{\leftarrow}, (t',p')_{\rightarrow}|\tau)}{(t-1) P((t, p)_{\leftarrow}).}
\end{split}
\end{equation}
We can then write the rate matrix as
\begin{equation}
\begin{split}
Q(t,p\rightarrow t',p') =\sum_{\tau=1}^{\min(t,t')-1} \frac{c_{p,t,p',t'}(\tau)}{2 \alpha_{p,t}(\tau+1)},
\end{split}
\end{equation}
where $c_{p,t,p',t'}$ is the proportion of nonmigrant (diploid) sites with joint ancestry $(p,t)$ and $(p',t'),$ which can be obtained using a recursive equation, as in equation \eqref{apt}. 
In the drift-less case, this reduces to 
$$Q(t,p\rightarrow t',p') =\sum_{\tau=1}^{\min(t,t')-1} \alpha_{p',t'}(\tau+1),$$
as obtained in Equation \eqref{dlrate}.

A case of particular interest is the pulse migration, with proportions $m$ and $1-m$ for populations 1 and 2, respectively. We then get $\alpha_1(\tau)=m$, and 
$$a_{p,t,p',t'}(\tau)=2 m (1-m) \left(1-\frac{1}{2N}\right)^{T-1-\tau}.$$

We can therefore calculate the transition probabilities, which are still proportional to the migration rates, but now exhibit a more complex time dependence: 

$$Q(i \rightarrow j\neq i)= m_j (2N-1) \left(1-\left(1-\frac{1}{2N}\right)^{T-2}\right).$$

The limit $N\rightarrow \infty$ yields the drift-less case
$$Q(i \rightarrow j\neq i)= m_j (T-2),$$
and the limit $T\rightarrow \infty$ reveals a linear dependence of the transition rate on the population size: 
$$Q(i \rightarrow j\neq i)= m_j (2N-1).$$ The infinite-time tract lengths are thus inversely proportional to the effective population size. 




\section{Appendix 2: Numerical estimation of tract length distribution}
\label{infinite}

In this section we describe how to obtain the expected distribution of tract lengths, given a set of Markov transition rates. 
A straightforward numerical solution strategy is to uniformize the transition matrix \cite{Stewart:1994p4967}. Uniformization uses the fact that self-transition probabilities can be adjusted without affecting the trajectory statistics, and in such a way that the total transition rate from each state is equal to the rate of the state with the highest transition rate, $Q_0$. Once all states have the same outgoing rate $Q_0$, the problem can be decomposed in two steps; a discrete calculation of the number of transitions in a given excursion, and a calculation of the trajectory lengths given the number of transitions. 

In the first step, we establish the distribution $\{b_n\}_{n=1,\ldots,\infty}$ of the number of steps spent in tracts of a given ancestry $p$, which is a standard discrete Markov excursion problem. In principle, the number of steps can be arbitrarily large, but the probability of very long tracts decays rapidly, and after a certain number of steps the expected length of the excursion is more than the chromosome length. We therefore calculate $\{b_n\}_{n=1,\ldots,\Lambda}$  up to a cutoff $\Lambda$, such that $\sum_{i=1}^{\Lambda} b_i \simeq 1$ (we usually also choose $\Lambda$ such that $\Lambda Q_0> L$, the length of a chromosome. To ensure a proper probability distribution, we then set $b_{\Lambda+1}=1-\sum_{i=1}^{\Lambda} b_i$. There are many ways to obtain the $b_n$. For our purposes, we have found it convenient to evolve the state vector by repeated multiplication with a transition matrix modified to have a single, absorbing state corresponding to the non-$p$ ancestries, and recording the amount of absorbed probability per multiplication.
 
The second step is straightforward since the length of the trajectories with $k$ steps follows the Erlang distribution 
 $$E_{k,Q_0}(x)=\frac{Q_0^k x^{k-1} e^{-Q_0 x} }{(k-1)!},$$  
 leading to the following expression for the tract length distribution: 

\begin{equation}
\phi(x)\simeq\sum_{k=1}^{\Lambda+1} b_k E_{k,Q_0}(x).
\label{ErlangExp}
\end{equation}


\subsection{End effects}
\label{finite}
Ancestry tract length distributions obtained in the infinite-chromosome limit may not be appropriate for finite genomes, particularly if many tracts have a length comparable to the chromosome length. For example, predicted tracts may be longer than the full chromosome length $L$, and these will not be observed.  To model the tract length distribution on a finite chromosome, we consider a general tract length distribution $\phi(x)$ on an infinite chromosome, and ask for the distribution of tract lengths observed in a given window of length $L$. To this end, we first calculate the probability that the intersection of a tract of length $x_0$ and a window of length $L$ has length $x$. The probability $P(I)$ that a tract of length $x_0$ intersects the window of length $L$ is proportional to $x_0+L$. Given $I$, and assuming that $x_0<L$, the probability that the intersection is of length $x$ is
\begin{equation}
P(x|x_0\leq L,I)=\frac{2}{x_0+L} \Theta(x_0-x)+\left(1-\frac{2 x_0}{x_0+L} \right)\delta(x-x_0),
\end{equation}
with $\Theta$ the Heaviside function and $\delta$ Dirac's delta function. 

The result for $x_0>L$ can be obtained by the permutation $x_0\leftrightarrow L$, so that
\begin{equation}
P(x|x_0\geq L,I)=\frac{2}{x_0+L} \Theta(\eta-x)+\left(1-\frac{2 \eta}{x_0+L} \right)\delta(x-\eta)
\end{equation}
with $\eta=\min(x_0,L).$ This yields
\begin{equation}
\begin{split}
P(x|x_0)\propto P(x|x_0,I)*(L+x_0).
\end{split}
\end{equation}
As a result, we can write the expected new tract distribution, ranging from 0 to L, as
\begin{equation}
\begin{split}
\phi'(x)\propto &2 \int_{x}^\infty dx_0 \phi(x_0)+ (L-x)\phi(x)\\&+\delta(L-x) \int_L^\infty (x_0-L)\phi(x_0) dx_0.
\end{split}
\end{equation}
The first term corresponds to the tracts that contact the edges of the window, the second term describes tracts that are strictly included in the window, whereas the third term describes all tracts that span the full window. Note that the `edge' tracts therefore generally have a different length distribution compared to the `inner' tracts, and that as $L$ goes to infinity, the second term dominates and is proportional to $\phi(x)$.  

The normalizing factor is
\begin{equation}
Z=L+\int_0^\infty x \phi(x) dx,
\end{equation}
if all tracts are taken into account, and 
\begin{equation}
Z^{<L}=L \left(2-\int_0^L \phi(x) dx\right)+\int_0^L x \phi(x) dx,
\end{equation}
if only tracts with length $x<L$ are considered. Finally, if only tracts of length greater than $C$ are considered, we have 
$$Z_C=\left(L-2C\right)\int_C^\infty \phi(x)+\int_C^\infty x \phi(x)$$
and 
$$Z_C^{<L}=\left(L-2C\right)\int_C^\infty \phi(x)+L\int_L^\infty \phi(x) dx+ \int_C^L x \phi(x).$$

We now apply these results to the tract length distributions from Equation \eqref{ErlangExp}. First, we note the Erlang distribution is related to the generalized incomplete Gamma function by:
\begin{equation}
\begin{split}
\int_{x_1}^{x_2} dx'E_{k,T}(x')&= \frac{\Gamma(T x_1,Tx_2, k)}{(k-1)!},\\
\int_{x_1}^{x_2} dx' x'E_{k,T}(x')&=\frac{\Gamma(T x_1,T x_2, 1+k)}{T (k-1)!}.
\end{split}
\end{equation}
This way, using our series expansion \eqref{ErlangExp}, everything can be calculated in terms of gamma functions. For example,
$$Z=L+\sum_{i=1}^{\Lambda+1} b_ik/T $$
and 
$$Z_C=Z-2C+(2C-L) \sum_{k=1}^{\Lambda+1} b_k \frac{\Gamma(0,TC,k)}{(k-1)!}- \sum_{k=1}^{\Lambda+1} b_k \frac{\Gamma(0,TC,k+1)}{T (k-1)!}.$$
We can thus write separately the probabilities of having inside, edge, or full tracts of various lengths:
\begin{equation}
\begin{split}
\phi_i(x)&=\frac{(L-x)}{Z}  \sum_{k=1}^{\Lambda+1} b_k E_{k,T}(x)\\
\phi_e(x)&=\frac{2}{Z}\sum_{k=1}^{\Lambda+1} b_k \frac{\Gamma(Tx,\infty,k)}{(k-1)!}\\
\phi_f(x)&=\frac{\delta(L-x)}{Z}\\&\times \sum_{k=1}^{\Lambda+1} b_k  \frac{ L \Gamma(T L,\infty, k)+\Gamma(T L,\infty, k+1)}{(k-1)!}.
\end{split}
\end{equation}

\section{Appendix 3: Ancestry variance in the absence of drift}

\subsection{Ancestry variance under a Markov model of ancestry}
\label{markovvariance}

We consider in this section the assortment variance, in the absence of drift, where ancestry in two individuals  is modeled as independent realizations of a two-state Markov process. Let the Markov states representing ancestry be labeled by $k=1,2$ with rates $q_1$ and $q_2$ out of states 1 and 2, respectively. The generalization to multi-state Markov processes is discussed below.
We first consider a single chromosome of length $L$, and are interested in the variance in $X$, the length of this chromosome covered in state $k=1$. We have $X=\int_0^L dx \psi_k(x)$, with $\psi_k(x)$ the indicator function of state $k$ at position $x$ along the genome. Changing the order of the expectation and the integrals, we have
\begin{equation}
\begin{split}
\mathbb{E}\left[ X^2\right]-\mathbb{E}\left[ X\right]^2= \int_0^L \int_0^L dxdy & \mathbb{E}\left[ \psi_k(x) \psi_k(y) \right]\\&-\mathbb{E}\left[ \psi_k(x) \right]\mathbb{E}\left[ \psi_k(y) \right]. 
\end{split}
\end{equation}
%
All these expectations are independent of the position along the chromosome. We therefore have
\begin{equation}
\label{x2x2}
\mathbb{E}\left[ X^2\right]-\mathbb{E}\left[ X\right]^2= \int_0^L \int_0^L dxdy  \alpha_k \left(P(y|x)-\alpha_k\right),
\end{equation}
with $\alpha_k=q_{1-k}/(q_{1}+q_{2})$ and $P_k(y|x)$ is the probability that $y$ is in ancestry $k$ given that $x$ is in ancestry $k$. In a Markov process, $P(y|x)=(1-\alpha_k) e^{-(q_{2}+q_{1})|x-y|}+\alpha_k.$ The integral yields
 $$\frac{\mathbb{E}\left[ X^2\right]-\mathbb{E}\left[ X\right]^2}{L^2}= \frac{2 \alpha_k (1-\alpha_k)}{(q_1+q_2)L}\left(1-\frac{1-e^{-(q_1+q_2)L }}{(q_1+q_2)L}\right).$$
In the absence of drift, $q_1+q_2=T-2$, and in the limit $(q_1+q_2)L>>0$, we recover our estimate: 
 $$\frac{\mathbb{E}\left[ X^2\right]-\mathbb{E}\left[ X\right]^2}{L^2}\simeq \frac{2 \alpha_p (1-\alpha_p)}{(T-2)L}.$$

 \subsection{Assortment variance for non-constant migration}
 \label{markovvariancecont}
The generalization to arbitrary one-way migrations is straightforward, in the absence of drift.  We evaluate Equation \eqref{x2x2} by expanding on arrival times $s$ for ancestry $p$:
\begin{equation}
\label{x2x2nc}
\mathbb{E}\left[ X^2\right]-\mathbb{E}\left[ X\right]^2= \int_0^L \int_0^L dx dy   \left(\sum_s P(y_p|x_{(p,s)}) \alpha_{p,s} -\alpha_p \alpha_p\right).
\end{equation}
 
The probability that $y$ is in the ancestry $p$, given that $x$ is in state $(p,s)$, can be written as $P(y_p|x_{(p,s)})  =\sum_{i,\nu} a^s_i e^{\kappa_i r}  v_{i\nu}\rho_{\nu p}$, where $r$ is the distance between $x$ and $y$ in Morgans, $\nu$ represents a Markov state $(p',s')$, $\rho_{\nu p}$ is the indicator that $p'=p$ and $(\kappa_i, v_{i\nu})$ are the eigenvalues and eigenvectors of the transition matrix $Q_{\nu\nu'}$. To obtain the $a^s_i$, we set $\sum_i a^s_i  v_{i\nu}=\delta_{\nu,(p,s)}$. For computational efficiency, we can first perform the sum over $s$ in equation \eqref{x2x2nc} and solve only once for $a_i=\sum_s a^s_i \frac{\alpha_{p,s}}{\alpha_p}$. Assuming that the Markov chain has a unique stationary distribution (which is the case if the number of generation is finite and last-generation migrants are not allowed), there is a unique $i_0$ with $\kappa_{i_0}=0$. The corresponding term cancels out in \eqref{x2x2nc}, so that we can finally write
\begin{equation}
\begin{split}
\mathbb{E}\left[ X^2\right]-\mathbb{E}\left[ X\right]^2&= \alpha_p \sum_{i\neq i_0,s} a_i v_{is}\rho_{sp} \int_0^L \int_0^L dx dy e^{\kappa_i r},\\
&= \alpha_p \sum_{i\neq i_0,s} a_i v_{is}\rho_{sp} \frac{(-1 + e^{L \kappa_i} - L \kappa_i)}{\kappa_i^2}.
\end{split}
\end{equation}

\subsection{Distinguishing the two components of the ancestry variance from inter-chromosomal variance}

\label{randomassort}

\begin{figure}
\scalebox{.32}{ \includegraphics{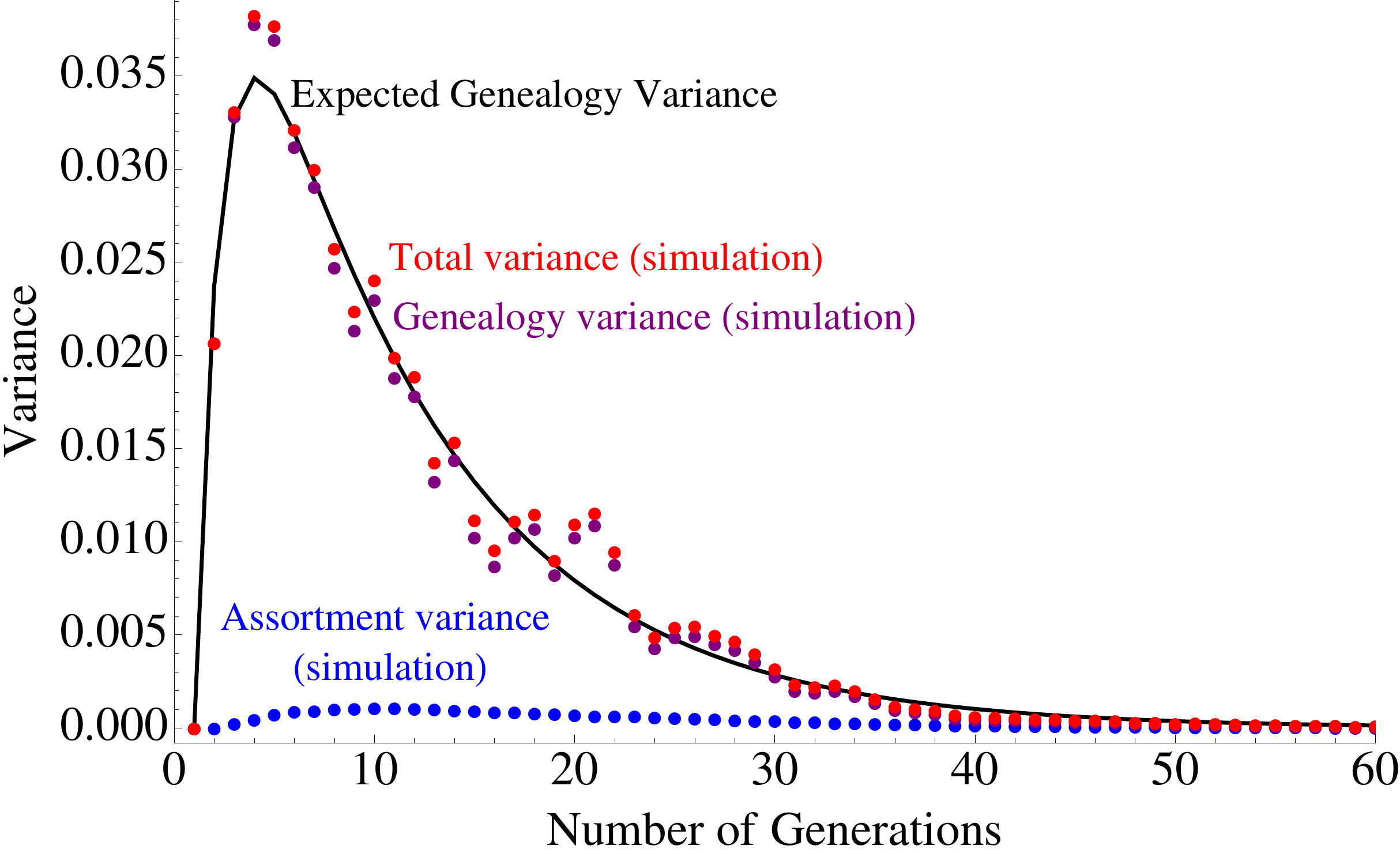}}
 \caption{\label{contvar} Time evolution of the variance for a population of 200 diploid individuals for a constant migration rate of $5\%$ starting at generation 1. As the fraction of genetic ancestry originating from the migrant populations grows from 0 to 1, the variance reaches a maximum before the migration frequency reaches $0.5$. Using the assumptions of Equation \eqref{varassumpt}, we decompose the observed variance (red dots) in a genealogy (purple) and an assortment (blue) contribution. As expected, the genealogy contribution dominates.}
 \end{figure}

We argued that, due to random chromosome assortment, the variance in ancestry between chromosomes is informative of the assortment variance. If all chromosomes had the same length, we could expect that the assortment variance in ancestry proportion across individuals would be proportional to the variance across chromosomes, and inversely proportional to the number of chromosomes per individual. However, the different chromosomes have different lengths, and to combine the information we need an idea of how ancestry variance depends on chromosome length. We assume that the assortment variance on ancestry is inversely proportional to the chromosome length in Morgans; in effect, we suppose that the number of independent ancestry observations is proportional to the chromosome length. The proportionality factor $\sigma_g$ depends on the pedigree, so that  
 \begin{equation}
 \label{varassumpt}
 \begin{split}
 \var(X^p|g)&=\var(\sum_i L_i X^p_i |g)/L^2\\
 &=\sum_i L_i^2\var( X^p_i |g)/L^2\\
 &\simeq \frac{\sigma^p_g}{L^2}\sum_i  L_i\\ 
 &=\frac{\sigma^p_g}{L}.
\end{split}
\end{equation}
 Furthermore, since we are interested in the average variance over all pedigrees, we get
 $$\mathbb{E}_g\left[\var(X^p|g)\right]=\frac{\mathbb{E}_g\left[\sigma^p_g\right]}{L} .$$
 We therefore wish to obtain an expression for $\mathbb{E}_g\left[\sigma^p_g\right]$ derived from the data. For each individual and each chromosome, we can obtain an estimate for this variance by comparing the ancestry proportion in that chromosome to the individual mean.  We can then obtain the best-fitting $\sigma^p_g$. An average over all sequenced individuals provides us with an estimate for $\mathbb{E}_g\left[\sigma^p_g\right]$. This procedure is used to decompose the simulated variances in Figure \ref{contvar}.
 
\end{document}